\documentclass[10pt]{iopart}

\usepackage{epstopdf}
\usepackage{graphicx}

\usepackage{iopams}  
\begin{document}

\title{Acoustic Measurements of the Stripe and the Bubble Quantum Hall Phases}

\author{M.E. Msall}
\address{Dept. of Physics \& Astronomy, Bowdoin College, Brunswick, ME 04011, USA} 
\address{ Max Planck Institute for Solid State Research, Heisenbergstrasse 1, D70569 Stuttgart, Germany}
\ead{mmsall@bowdoin.edu}
\author{W. Dietsche}
\address{ Max Planck Institute for Solid State Research, Heisenbergstrasse 1, D70569 Stuttgart, Germany}
\address{ Solid State Physics Laboratory, ETH Z\"urich, CH8093 Z\"urich, Switzerland}


\vspace{20pt}

\begin{abstract}
We launch surface acoustic waves (SAW) along both the $<110>$ and the  $<1\bar{1}0>$ directions of a Hall bar and measure the anisotropic conductivity in a high purity GaAs 2-D electron system in the Quantum Hall regime of the stripe and the bubble phases. In the anisotropic stripe phase, SAW propagating along the  "easy" $<110>$ direction sense a compressible behavior (finite resistance) which is seen in standard transport measurement only if current flows along the  "hard" $<1\bar{1}0>$ direction.  In the isotropic bubble phase, the SAW data show compressible behavior in both directions, in marked contrast to the incompressible quantum Hall behavior seen in transport measurements. These results challenge models that assume that both  the stripe and the bubble phase consist of incompressible domains and raise important questions about the role of domain boundaries in SAW propagation.  
\end{abstract}

%
%
%
\maketitle
%
\ioptwocol

\section{Introduction}
The physics of a two-dimensional electron system (2DES) in high magnetic fields is dominated by the interactions between the charges. Interactions lead to the correlated states of the fractional quantum Hall effect (FQHE)\cite{Tsui1982} but can also lead to a spatial modulation of the electron density into geometrical patterns or domains. Examples are the stripe \cite{Lilly1999a,Du1999} and the bubble phases \cite{Fogler1996} which have been found in higher Landau levels. A domain formation also occurs at the transition between different spin polarizations at certain fractional states. \cite{Eisenstein1990}   In the regime of the integer quantum Hall effect, interactions lead to quantum-Hall ferromagnetism \cite{Poortere2000,Jungwirth2001} when Landau levels with different spin-polarization cross. 
Standard resistance measurements have been used in the discovery of all of these domain structures. However, measuring the voltage drop at the edge of a sample usually does not give information about the resistivity in the bulk of the quantum-Hall system or at the length scale of the domain sizes.  A number of different techniques have been used to overcome this deficit, including electrostatic potential measurements by a SET at a fixed position, \cite{Verdene2007}  scanning photo luminescence, \cite{Hayakawa2013} and nuclear magnetic resonance.\cite{Stern2004}

Surface acoustic waves (SAW) are a powerful probe for studying low-dimensional electron states, \cite{Wixforth1986}  as demonstrated by their fundamental role in the discovery of the composite fermion. \cite{Willett1990} In this work, SAW provide a new window into the composition of the Quantum Hall stripe and bubble phases in the filling-factor region 4 to 5. The damping and the dispersion of the SAW sensitively depend upon the conductivity along the direct path between the exciting and detecting transducer.  Our experiment uses pairs of bifrequency transducers to measure the SAW propagation along straight line paths through the bulk of the 2DES along two principle axes at very low temperatures($ {\approx}$ 10 mK).  In the stripe phase, we find an enhanced resistivity if the SAW propagates along the  "easy" $<110>$ direction while standard resistance measurements show the the enhancement when current flows along the perpendicularly oriented "hard" $<1\bar{1}0>$ direction. In the bubble phase our new  SAW data yield a finite conductivity while standard transport data show the vanishing conductivity of a quantum Hall effect.

\section{The stripe and the bubble phase}

At the 9/2 filling of the Landau levels observed anisotropic magneto-resistances \cite{Lilly1999,Du1999} have been attributed to exchange interaction driven phase separation into stripe regions of alternating integer filling ($\nu$= 4 or 5). The anisotropy in magnetotransport at $\nu$ = 9/2 is  attributed to a charge density wave through which electrons in the partially occupied 2nd Landau level form domains.  The theoretical edges of these domains are considered to be  sharp \cite{Fradkin1999} which has, however, not yet been experimentally verified. The stripes are thought to extend along the $<110>$ axis, since measured resistances are much smaller if current is sent along this direction than if the current flows perpendicular to it. 

Commensurability oscillations in SAW-assisted microwave absorption suggest a stripe periodicity of 3.6 times the cyclotron radius $R_C\sim $ 100 nm. \cite{Kukushkin2011}   In that work, very short wavelength SAW ($< 1 {\mu}m$) were used in conjunction with optical excitation and detection to probe the anisotropic dispersion of the low energy excitations of the $\nu$ = 9/2 state. 

The observed resistance anisotropy disappears above  $\sim$ 150 mK  even though Hartree-Fock calculations estimate a relatively large 3.6 K energy for a $0.1 {\mu}m$ uniform charge density wave at $\nu$ = 9/2. \cite{Wexler2006,Lilly1999}  This discrepancy between formation energy and maximum temperatures for the observation of the anisotropy suggests that the presence of CDW domains alone is insufficient to account for the anisotropy of the stripe phases. Thus, it has been conjectured that a temperature activated disorder in the stripe pattern destroys the anisotropic behaviour.  As the temperature is raised, unbound disclinations disrupt the current paths and restore isotropic behaviour. \cite{Oga2001}  
 An alternate model attributes the transition to a Fermi surface instability where the Fermi surface changes from circular to elliptical.\cite{Oga2001}  Kee predicts that acoustic probes of the Fermi liquid will measure a maximum phase change when the SAW wave vector is along either the major or minor axis of the Fermi surface and that the sound wave is heavily attenuated (due to coupling to an overdamped Goldstone mode) along all other directions. \cite{Kee2003} 

In the present work, we explore the anisotropic conductivity of the 9/2 state at relatively long wavelengths ($\sim 10 {\mu}m$).  In this regime, the stripes represent a modulation of the carrier density with a periodicity much less than the SAW wavelength. Unlike lateral density modulations imposed by persistent photoconductivity \cite{Gerhardts1989} or patterned gates, \cite{Willett1997} the stripe phase is expected to have integer step changes in filling factor from one stripe to the next.  Since both filling factors cause conductivity minima, one expects SAW measurements to show a conductivity minimum, regardless of the direction of the SAW relative to the domain structure. This SAW behavior should not change at temperatures above 150 mK if the stripe disorder model is correct because the exact geometrical pattern of the stripes would not change the fact that the conductivity should be small everywhere. 

In additon to the stripe phase a bubble phase has been identified.\cite{Fogler1996} It is also thought to be formed by domains of different but integer filling factors in the 2DES, e.g., a matrix with filling factor 4 in which bubbles with filling factor 5 are embedded. Transport measurements show a reentrant quantum-Hall effect of 4 at a filling factor of about 4.2 to 4.3. Likewise a bubble phase is also observed between 4.7 and 4.8 which is formed by filling-4 bubbles in a filling-5 matrix which also shows a reentrant quantum-Hall effect but corresponding to filling factor 5. 

No anisotropy is expected for the bubble phase. Again, in a SAW experiment one would expect to observe only the behavior of vanishing conductivity because all of the 2DES should be in the quantum-Hall state, as concluded in SAW studies of another system where both types of domains are in the quantum Hall state. 
\cite{Dini2007}

The pinning of both the stripe and the bubble phase has been conjectured from the observation of a resonant mode in high-frequency-conductivity measurements.\cite{Sambandamurthy2008}  Our measurements, limited to just three frequencies between 200 and 700 MHz, cannot confirm a strong frequency dependence in the conductivity, but the potential connection of such modes to the SAW results is an important avenue of further study.

\section{SAW and Resistivity Measurements}

SAW are produced by interdigital transducers consisting of a large number of aluminum stripes which are alternatively connected to the two polarities of an rf-source. The in-plane electric field of the transducers couples to vertical strain in the GaAs by its anisotropic piezoelectricity. If the wavelength of a surface wave matches the period of the transducers then a SAW is launched. Thus the applied rf-frequency must be in resonance with the transducer or with one of its odd harmonics. Although the strain of the propagating SAW is mainly vertical, it is connected with an in-plane electric field pointing parallel to the propagation direction which leads to the coupling with the 2DES.

Changes in the conductivity influence the electron screening and therefore impact the magnitude and phase of the acoustic waves. The SAW  intensity drops according to $exp({-\Gamma}x)$ with $\Gamma$ being the attenuation per unit length in x-direction. The value of $\Gamma$ and the  velocity shift or, equivalently, the  phase shift of the SAW relative to a layer with infinite conductivity is approximated by: \cite{Simon1996}

\begin{equation}
\label{eq:Gamma}
\Gamma=\frac{{\alpha}k_{ref}( \sigma_{xx}/\sigma_{m)}}{1+( \sigma_{xx}/\sigma_{m})^2}
\end{equation} 

\begin{equation}
\label{eq:phase}
\frac{-\Delta\varphi}{\varphi_{ref}}=\frac{\Delta v}{v_{ref}}=\frac{\alpha}{1+( \sigma_{xx}/\sigma_{m})^2}
\end{equation}

where $v$ is the SAW wavespeed, $\varphi$ is the phase of the SAW, $k_{ref}$ the SAW wavenumber.   $\sigma_{xx}$ is the real part of the conductivity of the 2DES parallel to the propagation direction of the SAW and $\sigma_{m} {\approx} 3.7 \rm{x} 10^{-7}  ({\Omega}/{\Box})^{-1}$ is a conductivity constant at which the SAW absorption is maximal. $\alpha$ is the dimensionless piezoelectric coupling constant (${\approx} 6.4 \rm{x} 10^{-4}$) which depends upon the depth of the 2DES and the wavevector and is therefore frequently used as a fitting parameter.
The subscript $ref$ refers to the values for the SAW on GaAs in the absence of the 2DES. 

In a piezoelectric material, the strains of the surface acoustic wave are accompanied by electric field oscillations. The embedded 2-d conducting layer changes the screening response.  A low-conductivity layer behaves like a piezoelectric insulator with an increased wavespeed due to piezoelectric stiffening. A high-conductivity layer effectively screens the electric fields, lowering the wavespeed resulting in an equivalent relative change in the detected phase,  as in Eq. (\ref{eq:phase}).

The effect of a domain structure on the SAW has not yet been treated theoretically in detail. However, in a simple picture, a SAW may pass alternatively through domains of different conductivity and spatial extension while their effects on damping and wavespeed are integrated along the path of the SAW.  Earlier experiments have already tried to exploit these properties. Willett et al. produced a structure with an artificial  inhomogeneity by a gating superlattice.\cite{Willett1997} A damping of the wave was observed only near filling factor 1/2 and only if the wave travelled along the domain lines. Dini et al. investigated the spin transition of the 2/3 state which is connected with a huge longitudinal resistance increase (HLR) \cite{Kronmuller1998} but, surprisingly, did not find any sign of the domains in the simultaneously measured SAW propagation \cite{Dini2007}. Indications of inhomogeneities at the metal-insulator transition have been concluded from SAW measurements,\cite{Tracy2006} while the excitonic-condensate state at ${\nu}_{tot}=1$ appeared homogeneous for the SAW.\cite{Msall2013} For completeness, we mention that SAW has been used to move optically excited excitons in several systems.\cite{Rocke1997,Sogawa2001}

\begin{figure}
 \includegraphics[width=70mm]{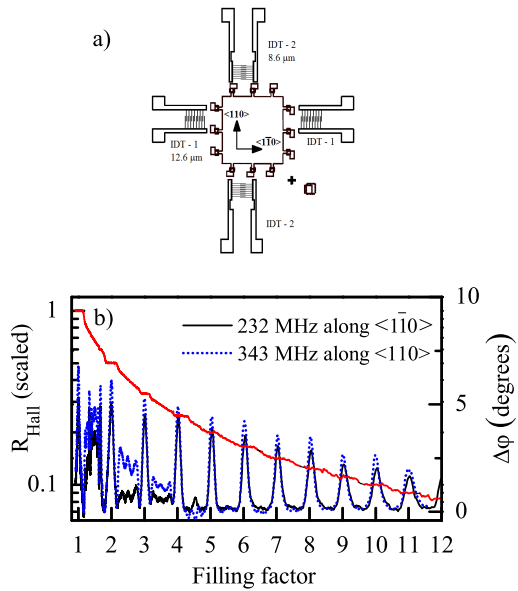}
 \centering
\caption{\label{Fig01}
\bf{(a)} Top view of the Hall bar contact pattern and SAW transducers.  The two sets of transducers (IDT-1 and IDT-2), which are connected pairwise in parallel, transmit SAW along perpendicular crystal axes at different frequencies.  \bf{(b)} Changes in SAW phase, measured in a time window centered on the first SAW arrivals, show the increase of SAW velocity coincident with Hall-resistance plateaus at integer filling factors.   
Note that the scaled $R_{Hall}$ axis is logarithmic. }
\end{figure}

In our experiments a 1 mm$^2$ 2DES is flanked by two sets of interdigital transducers (IDT-1 and IDT-2), as shown in Fig.~\ref{Fig01}.  The GaAs/AlGaAs heterostructures confine transport of the electrons to a 30 nm wide quantum well, 180 nm below the surface. The 2DES has an electron density of $2.7 \rm{x} 10^{11} cm^{-2}$. The shape of the 2DES, the "mesa" is similar to a van-der-Pauw geometry which allows standard resistance measurements in the two crystal directions.\cite{Goeres2007} 
The SAW are launched from the etched surface with negligible back reflection at the 120 nm high mesa edge because of the long wavelength. 

Temperatures in the 10 mK region are required to observe the 9/2 striped phase. This temperature requirement usually conflicts with the heat conduction of the coaxial cables that transmit the rf frequencies. Therefore, we thermally anchor the coaxial cable every 30 cm at the surrounding steel tube using Cu springs starting at the 300 K end. In addition, the lowest sections of the two cables are replaced by  NbTi cables which remain superconducting with small heat conduction in the magnetic fields used here. With these measures, the temperature of the mixture in the top loading dilution refrigerator is below 10 mK. The actual 2DES temperature is certainly higher. 

The  two transmitters and receivers are pairwise connected in parallel, so that we can operate two sets of transducers with just one pair of transmission lines. IDT-1 excites SAW centered at 232 MHz (or at the 3rd harmonic) that travel along the  $ <1\bar{1}0>$ crystalline direction; IDT-2 excites 343 MHz along the  $<110>$ direction (the 3rd harmonic of IDT-2 was not detected).  Each IDT transducer has a lateral width of 50 ${\mu}m$ and is aligned to one side of the center Hall contacts.  The SAW is generated and detected with an Agilent 5070 Network Analyzer. This instrument is equipped with a time resolution option enabling the collection of pulse-like data over a small frequency range \cite{Dini2007}. By changing the frequency window of the instrument we switch between the transducer resonances and the two crystal directions. To our knowledge, this bifrequency technique had not been used before for experiments in the quantum Hall regime. It should be mentioned that the bifrequency technique could easily be extended to many frequencies giving a plethora of experimental possibilities for high spatial resolution.  

The SAW experiments are accompanied by simultaneous standard electrical transport measurements using 8.4 Hz  currents of 5 nA and monitoring the voltages with Stanford 830 DSP Lock-in averagers. The  resistances are measured using central contacts on opposite edges for current and either two neighboring contacts on one side (longitudinal resistance) or two contacts on opposite sides (Hall resistance).

In  Fig.~\ref{Fig01}(b)  we show the velocity shift of the SAW, here expressed as the SAW phase shift, over a wide range of filling factors and for the two directions. The prominent peaks correspond to the vanishing conductivity at integer filling factors. Similar peaks are seen at fractional quantum Hall states below filling factor 2. Anisotropies between the different directions are visible between filling factors 2 and 3 and also between 3 and 4. However, the shape of the SAW curves is similar along both directions and these shapes are similar to those predicted by the (isotropic) longitudinal resistance data. Sample inhomogeneities which become more significant in the higher magnetic field regions could be the origin of the broad SAW anisotropy below filling factor 4. In the filling factor region between 4 and 5, however, the SAW behavior shows anisotropies which depend on the conductivity in a subtle way, described in the next section.

\begin{figure}
 \includegraphics[width=70mm]{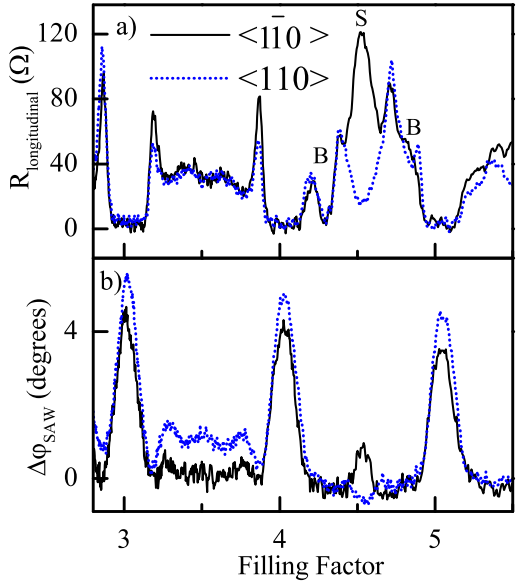}
 \centering
\caption{\label{Fig02}
\bf{(a) Longitudinal resistance measured along two different crystal directions. At filling factor 9/2 marked "S", the  resistance along the $<110>$ direction (color,dotted) is significantly lower than that measured along $<1\bar{1}0>$ (black) signalling the formation of the stripe phase.  The expected positions of the bubble phases are marked "B".  \bf{(b)} Changes in SAW phase associated with decreased conductivity are seen for SAW propagation along $<1\bar{1}0>$ (black)  at filling factor 9/2. The SAW phase is minimal along the $<110>$ direction at the same filling factor, indicating higher conductivity along this axis (color). }}
\end{figure}

\section{Results at 9/2}

The well-known anisotropy in the resistance of the 9/2 state is reproduced in our standard transport data as shown in Fig.~\ref{Fig02} (a), where the longitudinal resistance at 9/2 (marked with an "S") is significantly lower with current flowing along the "easy" $<110>$ propagation axis than along the "hard" $<1\bar{1}0>$ propagation axis.  Also shown in Fig.~\ref{Fig02} (b) are SAW phase shift data.  Note that a large phase shift corresponds to a smaller conductivity. A phase shift maximum is observed at $\nu$= 9/2  if the SAW is propagating along  $<1\bar{1}0>$. In the perpendicular direction no clear feature is visible. This observation is contrary to expectations. Assuming an integer filling factor of either 4 or 5 in the stripes one would expect a phase shift maximum comparable to the ones at $\nu$=4 and 5. Thus the SAW must sense a considerable finite conductivity on its path through the 2DES. 

Furthermore, the phase shift of the SAW is anisotropic with respect to the two crystal directions. A phase peak (conductance minimum) is observed if the SAW propagates along the  $<1\bar{1}0>$ direction. This is the direction in which the conventional transport measurement resulted in a resistance maximum. Thus a reduced conductance is sensed by the SAW if it propagates perpendicular to the stripes. This behavior can be understood if the proper inverse relationship between anisotropic resistivity and conductivity is taken into account, namely,

\begin{equation}
\label{eq:sigma1}
 \sigma_{xx}=\frac{\rho_{yy}}{Det{|\rho|}} .	
\end{equation}
				
where $|{\rho}|$ is the resistivity tensor. If, as it is in our case, the transverse resistivity is significantly greater than the longitudinal resistivity, then
\begin{equation}
\label{eq:sigma2}
\sigma_{xx} \approx \frac{\rho_{yy}}{\rho_{xy}^2} .
\end{equation}

Thus, following equation(\ref{eq:sigma2}), the anisotropy in longitudinal resistance at $\nu$= 9/2 should be mirrored in the SAW measurements along contrary axes, because it is $\sigma_{xx}$ which is relevant for the SAW, see Eq. (\ref{eq:phase}). And that is what is observed.

\begin{figure}
\includegraphics[width=70mm]{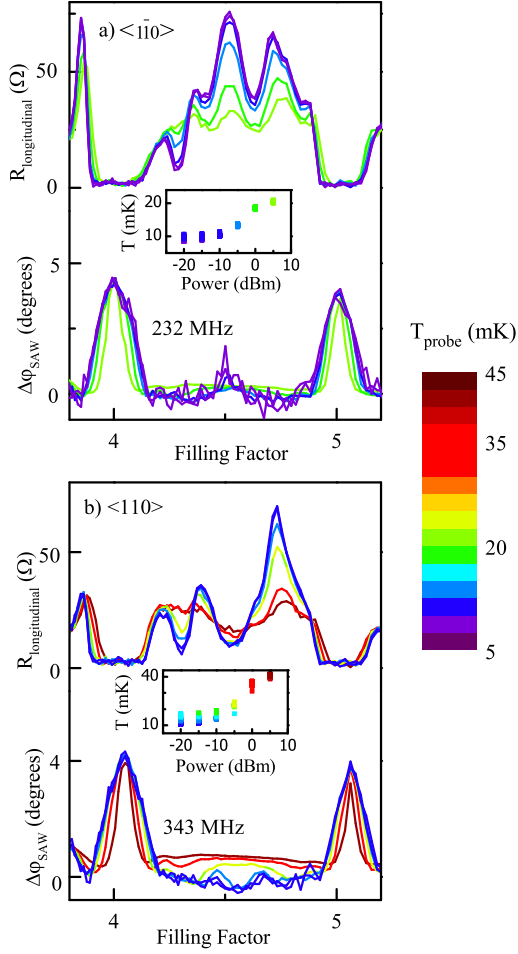}
\centering
\caption{\label{Fig03}
\bf {Longitudinal resistances and SAW phases as function of SAW power for the two directions. (a) $<1\bar{1}0>$ direction (232 MHz). (b) $<110>$ direction (343 MHz). An increase of the SAW power leads to a significant heating of the 2DES. The SAW phase peak observed at 9/2 along the $<1\bar{1}0>$ direction (232 MHz) is visible only at the lowest SAW power (a). The insets and the color scale show the reading of the temperature sensor mounted very near the sample.} }
\end{figure}

It is possible that the expected but missing observation of an incompressibility with the SAW is caused by a too high sample temperature, which could impede the formation of the ideal stripe phase.  The strength of the anisotropic response at $\nu$  = 9/2 is extremely sensitive to the applied SAW power, as shown in Fig.~\ref{Fig03}.  The system temperature also rises with increasing SAW power (insets, Fig.~\ref{Fig03}) but becomes significant only after the SAW power exceeds about -10dBm.  The observed SAW heating and the detected SAW magnitude are greater at 343 MHz which is probably caused by different transducer efficiencies. The SAW peak at 9/2 in the $<1\bar{1}0>$ direction disappears quickly with increasing power/temperature. The sluggish increase of the sensor temperature with SAW power indicates a weak thermal coupling between sample and sensor.  The sample temperature is likely to be considerably higher than the 10 mK shown by the sensor.

The anisotropic Fermi surface model predicts a difference in bulk properties that the SAW can directly sample.  It is predicted that the SAW phase change is maximum when the SAW wave vector is along either the major or minor axis of the Fermi surface and that the sound wave is heavily attenuated  along all other directions.  This is not visible in our data.  The SAW phase change is not a maximum at 9/2 for either propagation direction.  Along the $<1\bar{1}0>$ propagation axis, the SAW phase change at 9/2 is roughly 20\% of the observed phase changes at $\nu$ = 4 and 5; along the $<110>$ axis, the SAW phase does not change at all.  The measured SAW magnitudes (not shown) exhibit very similar sound attenuation at $\nu$ = 9/2 for both propagation directions, with no enhancement relative to other filling factors.

\section{SAW and the Bubble Phase}

Samples showing the stripe phase frequently also exhibit signatures of a "bubble phase" at filling factors of about 4.3 and 4.7.\cite{Fogler1996}  The bubbles are thought to consist of islands which differ by one in filling factor from the surrounding 2DES. These surmised bubble phases show the properties of the integer quantum-Hall effect in electrical transport measurements. No or only a weak anisotropy is observed at small currents.\cite{Goeres2007} In the present sample, a clear longitudinal resistance minimum at a filling factor of about 4.3 can be interpreted as being due to the bubble state, see Fig.~\ref{Fig02}, marked "B". The bubble state expected at a filling factor of about 4.7  is not well developed in our sample.

\begin{figure}
\includegraphics[width=70mm]{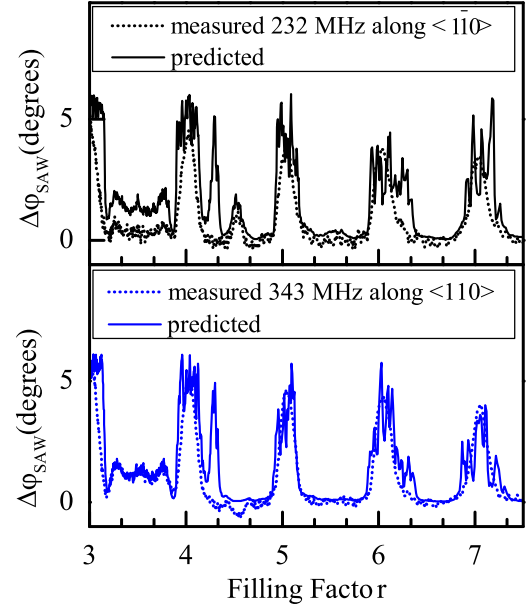}
 \centering
\caption{\label{Bubbles}
\bf{SAW phase shifts predicted from the measured anisotropic resistances using Eq. (\ref{eq:phase}) and (\ref{eq:sigma2}). They agree well with the measured ones except for a predicted peak near $\nu =4.3$ which is missed in the SAW data in both directions  This peak originates from the conductivity minimum attributed to a bubble phase. This indicates that the SAW passes substantial 2DES areas which have a finite conductivity. The predicted maximum SAW phase shifts appear noisy because the noise in the longitudinal resistance measurement is amplified by Eq. (\ref{eq:phase}). The curves are fitted to each other at the minimum and the maximum phase shifts.}}
\end{figure}

Fig.~\ref{Bubbles} compares the SAW phase shifts calculated from the transport data according to Eq. (\ref{eq:phase}), with the measured ones. Surprisingly, the distinct resistance minima  measured along both the hard and easy axes do not lead to any measurable phase shift of the SAW in either direction. In analogy to the interpretation of the SAW results at 9/2, we conclude that the SAW passes substantial regions with a finite conductance while the transport measurements are dominated by the 2DES  matrix with an integer filling factor.

\section{Conclusions}

The SAW measurements at $\nu=$ 9/2 show clear signatures of the stripe phase. The expected vanishing of the conductivity is, however, not reproduced in the SAW data. Rather,
 the anisotropy measured with standard resistance measurements is qualitatively reflected in the SAW, if one deduces the $\sigma_{xx}$ value relevant for the SAW from the resistivity $\rho_{yy}$ perpendicular to the respective SAW propagation direction. Thus, our results correspond to the established framework of the SAW absorption being determined by the local conductivity averaged over the SAW wavelength. One could visualize the action of the SAW as producing an electric field which forces charges to flow into the neighbouring stripe which affects the conductivity $\sigma_{xx}$.

There are similarities to the effects observed by Willett et al. \cite{Willett1997} in density modulated samples near filling factor $\nu$=1/2.  In their experiments, SAW phase peaks were observed when the SAW propagation direction was parallel to the lines of the periodic gates that controlled the density modulation.  A theoretical model for the $\nu$=1/2 case which assumes that the composite fermion scattering length is larger than the modulation period \cite{Oppen1998}  predicts a reduced resistivity perpendicular to the periodic gate lines even at small density modulations and, equivalently, to the  conductivity reduction required to explain the SAW data. A reduced resistance was indeed observed in later transport experiments when current was flowing perpendicular to the lines of an imposed superlattice.\cite{Smet1998}

In the regime of the bubble phase, the SAW propagation is dominated by a finite conductivity of the 2DES in both crystal direction. This is  contrary to the transport measurement which show a vanishing resistance in both directions. This seems to indicate that the bubbles themselves are not in an incompressible state but only the surrounding 2DES. Thus, the traditional resistance measurements are dominated by the incompressible 2DES to which the voltage probes are connected. The resistivity inside the bubbles can obviously not be probed by a standard transport experiment. 

There might be other explanations for the SAW behavior of the stripe and the bubble phase, e.g., that significantly lower sample temperature are needed to develop the theoretically anticipated situation. In such a case, future SAW experiments at even lower temperatures should show the expected incompressible behavior of both the stripe and the bubble phase. 

The sensitivity of the SAW to the stripe phase contrasts sharply with the behavior observed in the earlier work at the spin transition at $\nu$=2/3. There the domain structure leads to very large resistance anomalies in traditional transport but does not show up in the SAW propagation neither parallel nor perpendicular to the current flow direction.\cite{Dini2007, Dini2007b} It was argued that the resistance anomalies were caused by compressible domain walls with widths which are too thin for the SAW to sample. Our new results raise questions about this conclusion.  Alternatively, one might speculate that the observation of domains with SAW requires the presence of pinning modes which are present in the stripe and bubble phases while the lack of a SAW signature for spin domains could indicate a lack of pinning modes in that system.  
    
In summary, the bifrequency technique allows measurements which give new and tantalizing glimpses of the bulk conductivity, providing a strong complement to the information available in standard electrical measurements.  Our results challenge models that assume that both the stripe and bubble phases consist of incompressible domains, pointing instead to compressible bubbles in an incompressible matrix and the need for more detailed modeling of the effect of SAW fields at domain boundaries.

We thank Jurgen Smet for suggesting this experiment, Vladimir Umansky for providing the wafer, Benedikt Friess for his assistance on sample preparation, and Klaus von Klitzing for his constant interest and many discussions. Madeleine Msall also thanks Dale Syphers for his enlightening depiction of anisotropic local conductivity.

\bigskip

\bibliography {IoPAcousticAnisotropy}

\end{document}